\documentclass[sigconf]{acmart}
\usepackage{mathptmx} %
\usepackage{balance}

\newcommand{\ignore}[1]{}
\usepackage{fancyhdr}
\usepackage[normalem]{ulem}
\usepackage{url}
\usepackage{algorithm}
\usepackage[noend]{algpseudocode}
\usepackage{graphicx}
\graphicspath{{images/}}

\usepackage{hyperref}

\setcopyright{acmcopyright}

\copyrightyear{2019}
\acmYear{2019}
\setcopyright{acmcopyright}
\acmConference[ISCA '19]{The 46th Annual International Symposium on Computer
Architecture}{June 22--26, 2019}{Phoenix, AZ, USA}
\acmBooktitle{The 46th Annual International Symposium on Computer Architecture
(ISCA '19), June 22--26, 2019, Phoenix, AZ, USA}
\acmPrice{15.00}
\acmDOI{https://doi.org/10.1145/3307650.3322217}
\acmISBN{978-1-4503-6669-4/19/06}

\title{Bit-level Perceptron Prediction for Indirect Branches}

\author{Elba Garza}
\email{elba@tamu.edu}
\affiliation{Texas A\&M University and AMD}

\author{Samira Mirbagher-Ajorpaz}
\email{parisamir@tamu.edu}

\author{Tahsin Ahmad Khan}
\email{tahsinkhan@tamu.edu}
\affiliation{Texas A\&M University}

\author{Daniel A. Jim\'enez}
\email{djimenez@tamu.edu}
\affiliation{Texas A\&M University and \\ Barcelona Supercomputing Center}

\begin{document}
\onecolumn
\thispagestyle{empty}
\vspace*{1cm}
{\LARGE\bfseries Bit-level Perceptron Prediction for Indirect Branches\par}
\vspace{1cm}
{\Large Author manuscript\par}
\vspace{0.6cm}
Proceedings of the 46th International Symposium on Computer Architecture (ISCA '19), pp. 27-38 (2019)\par
\vspace{0.4cm}
DOI: \url{https://doi.org/10.1145/3307650.3322217}\par
\vfill
{\small This is the author manuscript. The definitive Version of Record was published by ACM in Proceedings of the 46th International Symposium on Computer Architecture (ISCA '19), pp. 27-38 (2019). DOI: \url{https://doi.org/10.1145/3307650.3322217}.\par}
\clearpage
\twocolumn
\setcounter{page}{1}

 \maketitle
 \thispagestyle{firstpage}
 \pagestyle{plain}

\section*{Abstract}
Modern software uses indirect branches for various purposes including, but not
limited to, virtual method dispatch and implementation of switch statements.  
Because an indirect branch's target address cannot be determined prior to execution,
high-performance processors depend on highly-accurate indirect branch
prediction techniques to mitigate control hazards.

This paper proposes a new indirect branch prediction scheme that predicts
target addresses at the bit level.  Using a series of perceptron-based
predictors, our predictor predicts individual branch target address bits based
on correlations within branch history.  Our evaluations show this new branch
target predictor is competitive with state-of-the-art branch target predictors
at an equivalent hardware budget. For instance, over a set of workloads
including SPEC and mobile applications, our predictor achieves a misprediction
rate of 0.183 mispredictions per 1000 instructions, compared with 0.193 for
the state-of-the-art ITTAGE predictor and 0.29 for a VPC-based indirect predictor.

    \section{Introduction} \label{sec:intro}

As object-oriented languages have become ubiquitous within software
application design, so have indirect branch instructions.  Early work by
Calder \& Grunwald~\cite{Calder1994Quantifying} shows that programs written in
object-oriented languages like C++ contain many indirect branch instructions,
on average 23 times as many compared to C programs, due to polymorphism.
Polymorphism~\cite{Cardelli1985Understanding} uses dynamically-dispatched
function calls implemented through indirect branches to support dynamic
subtyping.  Some object-oriented languages may generate a virtual function
call for \textit{every} polymorphic object
call~\cite{Srinivasan1996Technical}.

More recently, Kim et al.~\cite{Kim2007VPC} examined Windows applications on
real hardware, showing that 28\% of branch mispredictions were due to indirect
branches. For a some programs, indirect branches accounted for almost
half the mispredictions measured. Since indirect branch instructions incur the
same runtime misprediction penalty as conditional branch instructions, it is
imperative to have accurate indirect branch prediction mechanisms in place.

Indirect branches are not just necessary for virtual function calls.  Other
common program constructs like switch-case statements, jump tables, function
pointer calls, procedure returns, and interface calls also depend on indirect
branching. Indirect branch instructions are inherently more difficult to
predict than conditional instructions. Rather than simply predict
\textit{taken} or \textit{not taken}, an indirect branch's prediction requires
predicting a specific target address value. As an indirect branch instruction
may lead to any number of different known branch target values at runtime, its
prediction is more difficult than a simple decision between two values. Some
control-flow breaks caused by indirect branches are predictable.  Kaeli and
Emma's call/return stack~\cite{Kaeli1991Subroutine} is shown to accurately
predict procedure returns' target address values.

Several software-based strategies have been proposed to reduce the 
prevalence of indirect branch instructions in programs~\cite{Calder1994Reducing, 
Holzle1994Feedback, Grove1995RCPO, Aigner1996Eliminating, Porat1996Compiler}.
Such \textit{devirtualization} techniques lower the rate of indirect branch 
instructions by substituting them with one or more direct conditional 
branches~\cite{Ishizaki2000Study}. Unfortunately, these methods are costly, 
requiring static analysis of the whole program~\cite{Calder1994Reducing, 
Dean1995CHA}, extensive profiling~\cite{Holzle1994Feedback, Grove1995RCPO}, 
or a combination of both~\cite{Aigner1996Eliminating, Porat1996Compiler}.
Additionally, type inference for C++ programs using static analysis is already 
a known NP-hard problem~\cite{Pande1994Determination}. As such, indirect 
branch instructions are inevitable breaks in control flow that 
must be properly addressed. Section~\ref{subsec:rel:elim} talks about these 
software techniques in more depth.

An assortment of hardware-based techniques to predict indirect branch
instruction targets have been proposed~\cite{Lee1984Strategies,
Calder1994Reducing, Nair1995Correlation, Chang1997Target, Driesen1998Accurate,
Driesen1998Cascaded, Kalamatianos1998Compression, Driesen1999Multi,
Seznec2006TAGE, Kim2007VPC, Xie2011TAP}.  These predictors output a target
prediction at runtime based on branch history, {\em i.e.} target address
values from previous branch executions. The most well-known example of these
is the branch target buffer, or BTB~\cite{Lee1984Strategies}.  Many
hardware-based indirect branch predictors maintain target values in dedicated
storage~\cite{Nair1995Correlation, Chang1997Target, Driesen1998Accurate,
Driesen1998Cascaded, Kalamatianos1998Compression, Seznec2006TAGE}, which can
account for significant die area and lead to increased power consumption.
Other approaches take advantage of already-present hardware structures to
facilitate indirect prediction~\cite{Kim2007VPC, Xie2011TAP}.
Section~\ref{subsec:rel:ind} discusses these hardware-based prediction
strategies at length.

This paper introduces a new indirect branch prediction algorithm, Bit-Level
Perceptron-Based Indirect Branch Predictor (BLBP).  It uses branch history and
perceptron-based learning~\cite{Block1962Perceptron, Jimenez2001Perceptron} to
predict individual target value bits. The processor accesses a specialized
BTB-like structure to select the target that matches most closely at the bit
level among all observed targets of the branch.

The predictor is competitive with state-of-the-art indirect branch prediction 
schemes, reducing misprediction rates by 5\% over ITTAGE~\cite{Seznec2011ITTAGE} 
while maintaining the same hardware budget.

\begin{figure*}[htbp!]
  \centering
  \includegraphics[width=1.0\textwidth]{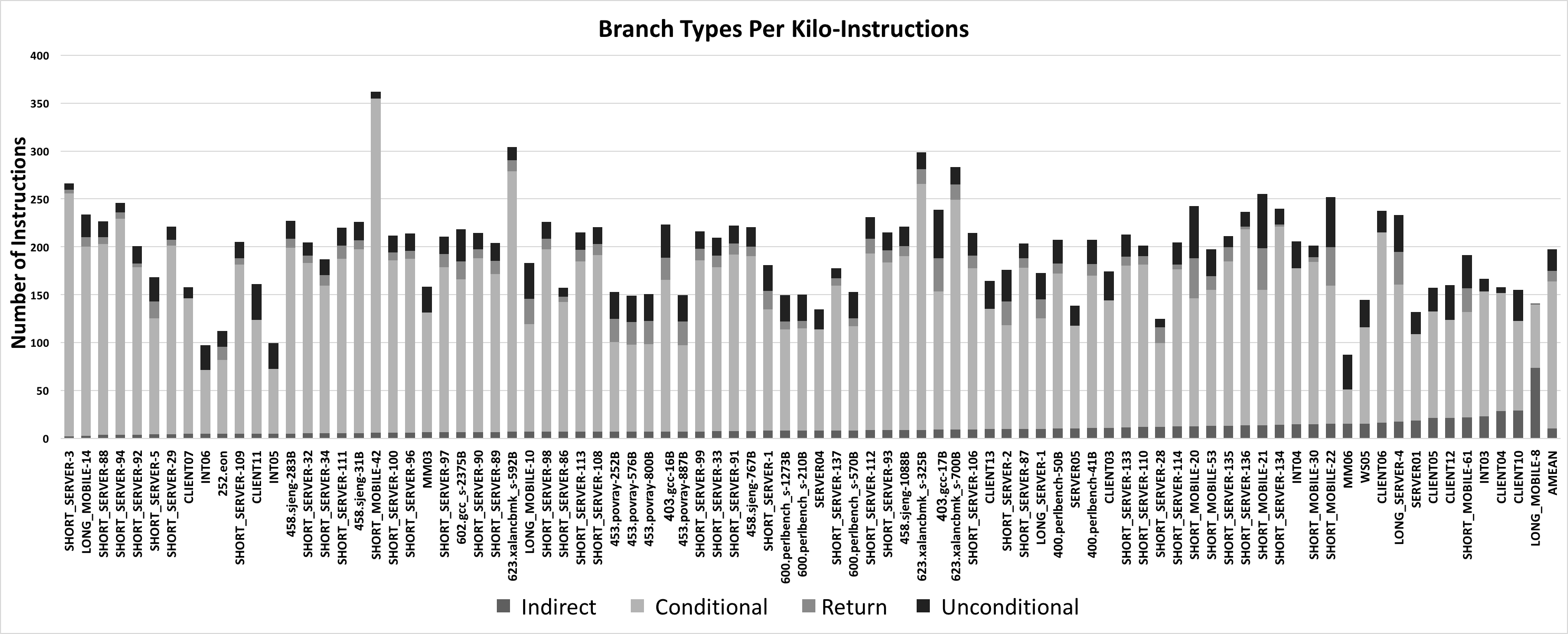}
  \caption{Breakdown of the prevalence of each branch type per kilo-instruction. Benchmarks are sorted by increasing prevalence of indirect branches.}
  \label{fig:types}
\end{figure*}

    \section{Related Work} \label{sec:related}

Control flow breaks due to indirect branch instructions can be mitigated by
two main strategies. The first strategy requires software-based schemes:
profiling and compiler optimizations can learn from programs to predict branch
targets, or minimize the frequency of indirect branches. The other strategy
depends on hardware mechanisms that predict branch target addresses
dynamically. The following two sections detail these strategies' related work.

\begin{figure}[ht]
  \centering
  \includegraphics[width=0.42\textwidth]{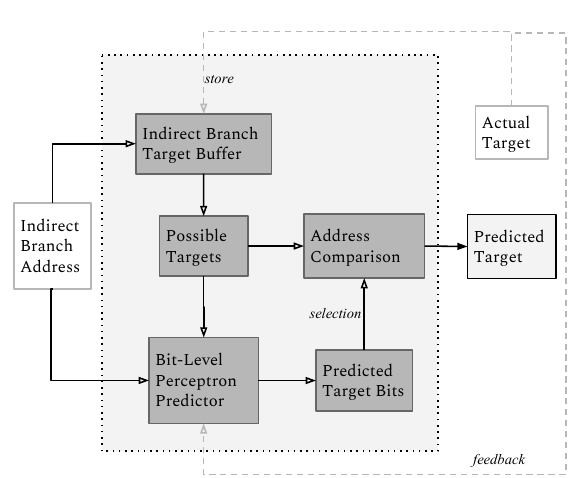}
  \caption{An overview of BLBP Target Address Prediction \& Selection}
  \label{fig:overview}
\end{figure}

\subsection{Indirect Branch Target Prediction Via Software}\label{subsec:rel:elim}

Wall proposed the first methods to statically predict indirect branches based
on profiling~\cite{Wall1991Limits}.  Calder \&
Grunwald~\cite{Calder1994Reducing} also used static profiling along with
compiler-driven optimizations and hardware modifications (see
Subsection~\ref{subsec:rel:ind}) to reduce C++ indirect function call
overhead.  Aigner \& H\"olzle proposed compiler optimizations that reduces the
frequency of virtual function calls~\cite{Holzle1994Feedback} using profile
driven type feedback and class hierarchy analysis~\cite{Dean1995CHA}. Porat et
al.~\cite{Porat1996Compiler} introduced similar inlining optimizations.  Roth
et al.~\cite{Roth1999Improving} devised a virtual function call prediction
technique which uses pre-computation to perform target address computation
before execution time.  Joao et al.~\cite{Joao2008Predication} proposed
Dynamic Indirect Jump Predication, using dynamic predication to handle
hard-to-predict indirect branches without prediction.  Farooq et
al.~\cite{Farooq2010VBBI} proposed Value-Based BTB Indexing, or VBBI
identifies hard-to-predict indirect branches through profiling and flags via
an ISA format augmentation, storing multiple targets for an indirect branch in
the BTB.  Compiler-Guided Value Pattern (CVP) prediction by Tan et
al.~\cite{Tan2012CVP} is similarly value-based and compiler-driven.

Software-based strategies are not free of hardware costs, however. Many works 
proposed above require supplemental hardware structures (even if minimal) and/or 
explicit ISA support~\cite{Calder1994Reducing, Joao2008Predication, 
Farooq2010VBBI, Tan2012CVP, Ghandour2014Leveraging}.

\subsection{Hardware Indirect Branch Prediction}\label{subsec:rel:ind}
 
The simplest target prediction hardware mechanism is the Branch Target Buffer,
or BTB~\cite{Lee1984Strategies}, using the branch address to index and cache a
branch's last-taken target address. The stored target address is later fetched
for target prediction. A BTB is sufficient for {\it monomorphic} branches
leading to just one target address, like direct branch instructions. However,
because indirect branches can be \textit{polymorphic}, or lead to different
addresses, the BTB's last-used prediction strategy is often
insufficient~\cite{Chang1997Target, Kalamatianos1998Compression}.  Calder \&
Grunwald improve this scheme using a 2-bit Branch Target
Buffer\cite{Calder1994Reducing} that replaces a target in the BTB only after
two consecutive mispredictions. However, BTB accuracy remains poor for
indirect branches.

Path-based history is also correlated with branch
behavior~\cite{Young1994Correlation, Young1995Comparative,
Nair1995Correlation}.  Unlike correlation with pattern
history~\cite{Yeh1991Two, Pan1992Improving, Yeh1993Comparison}, path-based
correlation uses the sequence of basic block addresses leading up to a branch.
Chang et al.~\cite{Chang1997Target} explored tracking both pattern and
path-based history in their indirect branch prediction scheme, the Target
Cache.

Driesen \& H\"olzle proposed Cascaded Predictors~\cite{Driesen1998Cascaded,
Driesen1999Multi} that are a hybrid of two target predictors, where the first
BTB-based predictor acts as a filter to catch easy-to-predict branches, while
hard-to-predict branches are predicted by a second two-level adaptive
predictor. The multi-stage predictor~\cite{Driesen1999Multi} generalizes the
cascaded predictor.

Kalamatianos \& Kaeli apply prediction by Partial Matching
(PPM)~\cite{Chen1996Compression} to indirect branch
prediction~\cite{Kalamatianos1998Compression}.  Seznec's TAGE, ITTAGE, and
COTTAGE predictors~\cite{Seznec2006TAGE} take inspiration from a PPM-like
predictor~\cite{Michaud2005PPM}. The TAGE predictor predicts conditional
branch directions while the ITTAGE predictor predicts indirect branch targets.
The COTTAGE predictor incorporates both a TAGE and ITTAGE predictor in one to
predict both branch directions and targets.  The predictors use geometric
history lengths~\cite{Seznec2005OGEHL} to index several partially-tagged
predicting tables.

Kim et al.'s Virtual Program Counter (VPC) predictor uses a hardware-based
devirtualization technique to predict indirect branch targets.  VPC is based
on the idea that a polymorphic branch instruction with $T$ different known
targets can be thought of as a series of $T$ individual direct branch
instructions.  The predictor attempts to ``devirtualize'' indirect branch
instructions in hardware to predict which of the $T$ targets is the correct
target output.  VPC's main advantage is that, rather than use dedicated
hardware, it reuses the existing conditional branch predictor and BTB.  The
conditional branch predictor is queried and the BTB accessed using a sequence
of ``virtual PCs'' corresponding to at least the $T$ targets of the indirect
branch.  The first virtual PC to output a \textit{taken} prediction (if any)
has its BTB target value output as the target prediction.  Since the virtual
PCs are are visited in series, the worst case scenario of no taken target may
result in many wasted cycles. Sorting the targets by frequency allows the
average latency to be low. A disadvantage of the VPC approach is that the
conditional branch predictor and indirect branch predictor rely on one central
prediction component that cannot be specialized to either task, but must be
tuned to give good general overall performance.

TAP prediction~\cite{Xie2011TAP}, predicts the \textit{address of a BTB entry}
that contains the predicted target value.  TAP predicts the BTB entry's
address bit-by-bit using several small branch predictors (called
sub-predictors) fashioned from the main conditional branch predictor.  The
idea is similar to BLBP, but TAP predicts the address of a BTB entry, not the
target address itself. Note that TAP prediction modifies hardware to allow
multiple different BTB entries for a single indirect branch address.

\begin{figure*}[ht] 
 \centering
 \includegraphics[width=\textwidth]{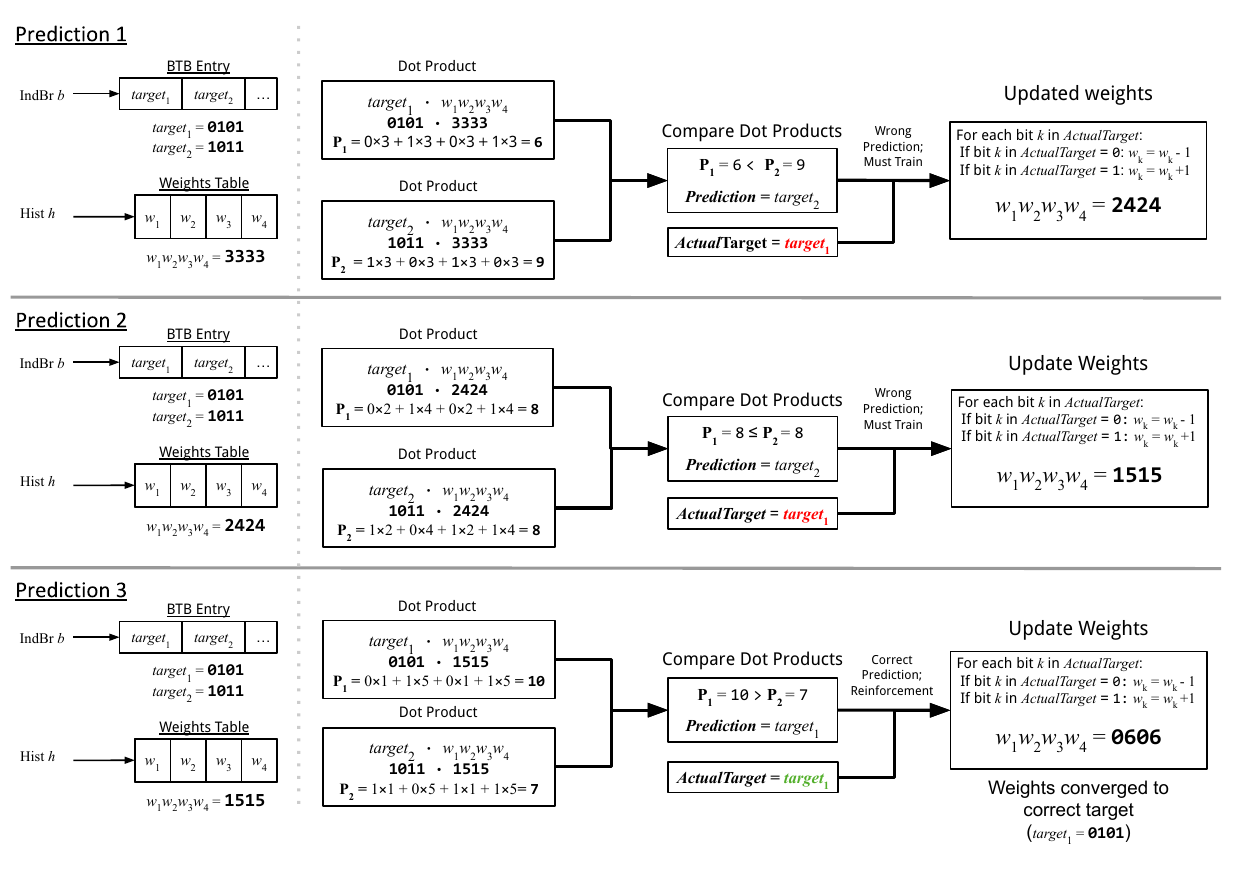}
 \caption{Through perceptron learning, BLBP trains weights to converge into bits 
 of the correct target. Normalizing the trained weight vector $(0,6,0,6)$ 
 would equal $(0,1,0,1)$, which is equal to the correct target bits in the 
 above example.}
 \label{fig:example}
\end{figure*}

\subsection{Perceptron-based Branch Prediction}\label{subsec:rel:perc}

Jim\'enez \& Lin~\cite{Jimenez2001Perceptron} first proposed the perceptron 
predictor, which bases its prediction scheme on the most basic neural-learning 
structure, the perceptron~\cite{Block1962Perceptron}. A single-layer perceptron 
is composed of a vector of integer weights $w_0 \dots w_n$. The output of a 
perceptron is computed as the dot product of the weighted vector with an input 
vector, $x_1 \dots x_n$, where $w_0$ behaves as the bias input. As such, the 
perceptron predictor maintains a table of weighted vectors rather than the more 
common saturating bits~\cite{Smith1981Strategies, Lee1984Strategies, Yeh1991Two, 
McFarling1993Combining}.
    
To predict a branch outcome, a weighted vector is chosen by indexing the table
using the branch's PC address.  The dot product is computed with the chosen
weighted vector and the global history register (of binary branch outcome
results) as input vector. If the dot product result is positive, a \textit{taken} 
prediction is output; otherwise, a \textit{not taken} output is given.

Due to the perceptron predictor's long prediction latency, Jim\'enez later
proposed the path-based neural predictor~\cite{Jimenez2003Path}.  The path-based 
neural predictor increases prediction accuracy while diminishing its prediction 
delay.

Piecewise linear branch predictor~\cite{Jimenez2005Piecewise} improves on the 
original perceptron predictor by learning previously-unlearnable linearly 
inseparable branches.   

Tarjan \& Skadron's hashed perceptron predictor~\cite{Tarjan2005Hashed}
improves upon the path-based predictor to further reduce prediction latency
and increase prediction accuracy while maintaining less physical state.

    \begin{figure}[ht!] 
 \centering
 \includegraphics[width=2.9in]{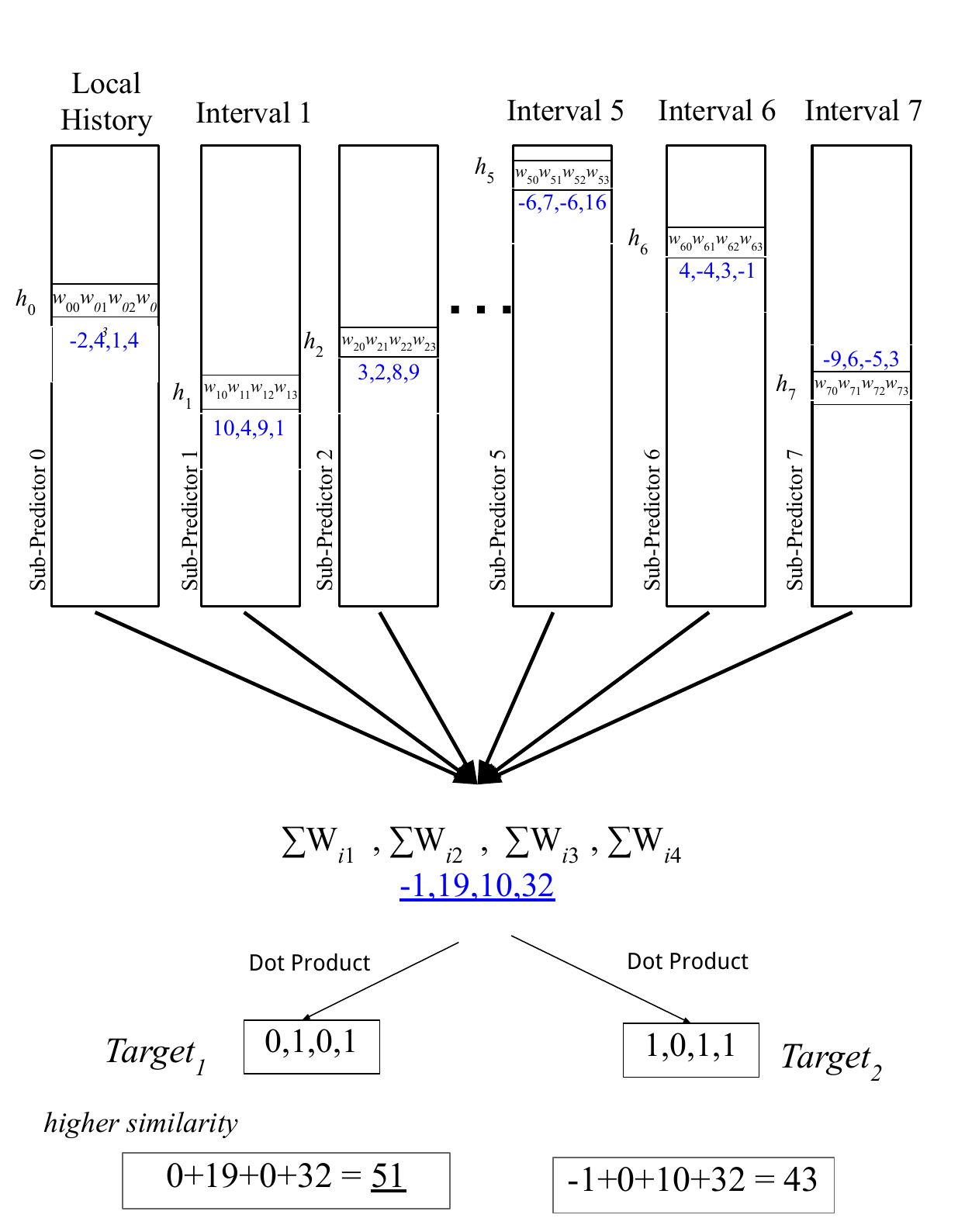} %
 \caption{BLBP makes a prediction by aggregating the likelihood for two example 
 targets from eight sub-predictors. Each sub-predictor is trained with 
 different history lengths or local history.}
 \label{fig:example-subpreds}
\end{figure}

\section{Bit-Level
Perceptron-Based Indirect Branch Predictor}\label{sec:snip}

BLBP is based on the Scaled Neural Indirect Prediction (SNIP) 
predictor~\cite{Jimenez2011SNIP}, which originally presented at a branch 
prediction workshop without published proceedings. The SNIP predictor proposes 
a novel indirect branch predictor that predicts the individual lower-order bits 
of target addresses, then finding the closest matching address in an indirect 
branch target buffer.
    
Our extension to SNIP, BLBP, improves accuracy beyond the state-of-the-art and
provides a path to a feasible implementation, greatly reducing the number of
SRAM arrays that would be needed for a practical implementation from 44 to 8.
The individual bits are predicted with neural predictors trained with perceptron 
learning.

\subsection{Indirect Branch Target Buffer(IBTB)}

BLBP uses a 64-way set-associative indirect branch target buffer (IBTB)
indexed by indirect branch PC to store up to 64 indirect branch targets.
Ideally there will be one set of targets per branch, but in practice many
branches may index the same set.  Thus each IBTB entry is tagged with 9 bits
from the branch PC.  The IBTB is managed with re-reference interval
prediction~\cite{Jaleel2010High} replacement. Figure~\ref{fig:overview} gives 
an overview of the process for predicting a target with BLBP.  

\subsection{Perceptrons For Each Bit of the Target}

BLBP trains perceptrons the same way as a hashed perceptron~\cite{Tarjan2005Hashed}, 
but rather than training a single weight for each target, a $K$-length vector 
of weights is trained, one for each bit of the target address that is being 
predicted.
  
Training by updating the $K$-length vector is shown in Algorithm~\ref{alg:update} 
and described in detail with example in Section~\ref{subsec:snip:update}.

BLBP has eight history features indexing tables we call {\em sub-predictors}.
Each sub-predictor predicts the values for each possible target that reflects
the confidence of that sub-predictor that the corresponding bit in the target
is 1.  For each bit in each possible target, BLBP adds the output of each
sub-predictor for that bit and build a vector ($y_{\mbox{out}}$). BLBP then
computes the similarity of each possible target with the vector. The target
with highest similarity will be selected as the prediction.
Figure~\ref{fig:example-subpreds} shows an example for 4 bit target
prediction. The first target has a higher similarity (Dot Product 51) with the
trained perceptron than the second target(Dot Product 43).

\subsection{Training For Multiple Histories} \label{subsec:snip:histories}
BLBP records the outcomes (\textit{taken} or \textit{not taken}) of recent
conditional branches into 630 bit global history($GHIST$).  Similar to hashed
perceptron learning~\cite{Tarjan2005Hashed}, BLBP uses different lengths and
sections of global history to train seven different sub-predictors.  BLBP also
records the outcomes(\textit{taken} or \textit{not taken}) of previous
iterations of the last 256 branches individually into 10 bit local histories.
BLBP indexes one sub-predictor with local history. More details on various
histories used in BLBP is explained in Section~\ref{subsec:snip:optimize}

\subsection{Predicting with Perceptrons}\label{subsec:snip:predict}   

\begin{algorithm}[thb!]
\caption{Predict Algorithm}
\label{alg:predict}
{\small
  \begin{algorithmic}[1]
    \Function{Predict}{\textit{pc}: \textbf{address}} : \textbf{integer vector}
      \For {$i = 1 \dots N$ \textbf{in parallel}}  %

    \State $j = \textrm{hash}(\textrm{GHIST}[1 \dots G_i])\textrm{ mod } M$  %

     \State $y_{\mbox{\small out}}[1 \dots K] \gets 0 $   
 \For {$k = 1 \dots K$}  %
          \State  $y_{\mbox{\small out}}[k] \gets y_{\mbox{\small out}}[k] + W[i][j][k]$
\EndFor   %
\EndFor  %

     \State $sum[1 \dots T] \gets 0 $   
     \State $max \gets 0 $   
\State $maxTarget \gets 0 $  
     \For {$t = 1 \dots T$ \textbf{in parallel}} %

 \For {$k = 1 \dots K$}  %
          \State  $sum[t] \gets \textit{sum[t]} + y_{\mbox{\small out}}[k] \times  \textit{target[t][k] } $
\EndFor   %

\If {\textit{sum[t]>max}}
\State $\textit{max} \gets \textit{sum[t]}$ 
\State $\textit{maxTarget} \gets \textit{t}$ 
\EndIf

\EndFor %

    \Return $\textit{maxTarget}$
    \EndFunction
  \end{algorithmic}
}
\end{algorithm}

Figure~\ref{fig:example-subpreds} shows how BLBP makes predictions by
aggregating the perceptrons from its eight sub-predictors and computing the
similarity with each possible target.  The first sub-predictor is trained with
local history. The next seven sub-predictors are trained with different history
lengths and history intervals explained in~\ref{subsec:snip:histories}.

Figure~\ref{fig:example} shows a simplified numeric example of the prediction
and training algorithm with only one sub-predictor.  For simplicity we show
only two targets. We also assume only 4 bits of targets to be predicted.

First the predictor uses the history to index the weights table and extract 4
weights, one weight for each target bit. Weights are between
-7 and 7. Here we assume each weight is equal to 3 initially.  Then the
predictor calculates the dot product of weights with the first targets bits.
The same 4 weights will be used to compute the dot product for the second
target.  Next, the dot products will be compared. The target with the maximum
dot product will be chosen as the predicted target.  In this example in the 
first prediction the output of the second target's dot product had a higher 
value than the dot product for the first target. Thus, the second target is 
chosen as the predicted target.

\subsection{Training Perceptrons}\label{subsec:snip:update}

\begin{algorithm}[H]
\caption{Update Algorithm}
\label{alg:update}
{\small
  \begin{algorithmic}[1]
    \Function{Update}{\textit{bits}: \textbf{address}, \textit{target}: \textbf{address}, \textit{suppress}: \textbf{address}}
      \For {\textit{k} = $ 1 \dots K$ \textbf{in parallel}}
        
        \If {\textit{suppress}$[k]$}
          \State continue
        \EndIf
        
        \State \textit{a} $\gets | \textrm{\textit{bits}}[k] |$
        \State \textit{correct} $\gets (\textit{target}[k] = (\textrm{\textit{bits}}[k] \geq 0))$
        \State adaptive\_training(\textit{correct}, $a$, $k$)
        
        \If {($!correct$ \textbf{or} $(a < theta[k]))$}
          \For {$i = 1 \dots N$}
            \State $j = \textrm{hash}(\textrm{GHIST}[1 \dots G_i])\textrm{ mod } M$
            \If { $(\textrm{\textit{target}}[k])$ } 
              \State increment\_unless\_at\_max(W[i][j][k])
            \Else
              \State decrement\_unless\_at\_min(W[i][j][k])
            \EndIf
          \EndFor
        \EndIf
      \EndFor
    \EndFunction
  \end{algorithmic}
}
\end{algorithm}

To show how the predictor learns from a misprediction, we continue the same 
example from Figure~\ref{fig:example}. In the first prediction, we saw how the 
second target was chosen as the predicted target through the dot product and 
comparison. Now we assume the first target was the actual target. Then each of 
the 4 weights that were used to make this prediction is adjusted with following 
rule:\\ 

For each weight used to predict bit $i$ of the actual target,   

\begin{enumerate}
 \itemsep0em 
 \item if $i$ is one then the weight is decremented,
 \item if $i$ is zero then the weight is incremented. 
\end{enumerate} 
  
Incrementing and decrementing weights will be saturated at the maximum or minimum 
values for the width of the weight. See Algorithm~\ref{alg:update} for more 
details. 
In our example after the first prediction, the actual target's bits are 
(0,1,0,1) so the weights are update from (3,3,3,3) to (2,4,2,4). 
Figure~\ref{fig:example} shows how after the second misprediction the weights 
are updated from (2,4,2,4) to (1,5,1,5), leading to a correct prediction. 
Assuming that the dot product for the actual target (here 10 for the first 
target) does not exceed the training threshold the weights will be updated after 
a correct prediction too. Figure~\ref{fig:example} shows in our example the four 
weights after the correct prediction are update to (0,6,0,6). Normalizing the 
trained weight vector (0,6,0,6) would be (0,1,0,1) which is equal to the correct 
target bits. It is an example of how BLBP trains weights to converge to bits of 
correct targets through perceptron learning. 

    \subsection{Optimization Techniques}\label{subsec:snip:optimize}

We use several optimizations to improve the accuracy of the predictor:

\paragraph{Local History} 
The first weights vector predicting a given bit is indexed with a hash of
{\em local history}. In conditional branch prediction, local history is a
shift register with the outcome of previous iterations of the branch being
predicted. For our predictor, we use shift registers with bit 3 of the
target.

\paragraph{History Intervals} The remaining weights vectors are indexed
using a hash of global history according to history intervals inspired by
the interval capability in multiperspective perceptron
prediction~\cite{JimenezCBP5} and strided sampling hashed perceptron
prediction~\cite{JimenezSSHP}. The intervals are tuned to achieve low
MPKI. The following 7 global history intervals were used for the 7
non-local vectors: (0,13), (1,33), (23,49), (44,85), (77,149), (159,270),
(252,630). For example, the second non-local vector was indexed by a hash
of global history outcomes from position 1 in the global history through
position 33. The tuned intervals were found by starting with geometric
histories~\cite{Seznec2005OGEHL} and improving with hill-climbing,
changing the start or end of an interval randomly and keeping the change
if it improved MPKI.

\begin{figure}[htbp!]
  \centering
  \includegraphics[width=0.45\textwidth]{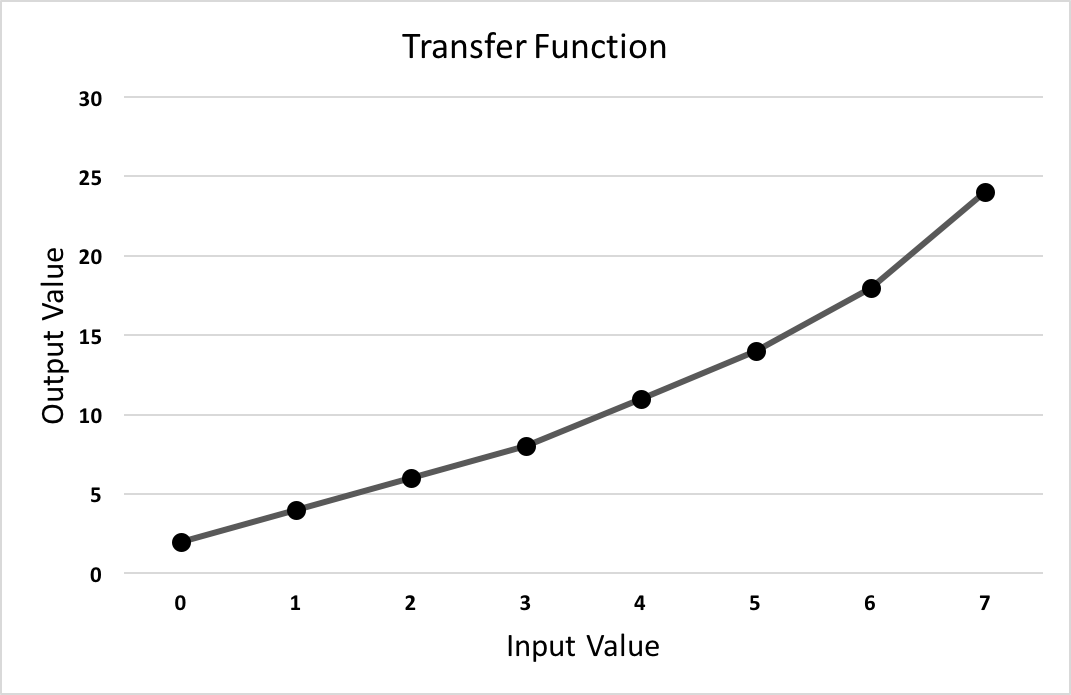}
  \caption{A transfer function amplifies the effect of higher
    weights and diminishes the effect of lower weights. }
  \label{fig:xfr}
\end{figure}

\paragraph{Transfer Functions} As with multiperspective perceptron
prediction~\cite{JimenezCBP5}, we find that applying a non-linear transfer 
function to the weights before summation boosts accuracy. From previous
work, we know  that a convex function can amplify the effect of higher
weights and diminishes the effect of lower weights, allowing the limited range
of weights values to more precisely model the predicted bits.
Figure~\ref{fig:xfr} illustrates the transfer function we developed empirically.

\paragraph{Adaptive Threshold Training} As with O-GEHL and previous perceptron
predictors, we use Seznec's adaptive threshold training algorithm to adjust
the threshold for training such that the number of instances of training on
correct predictions roughly equals the number of
mispredictions~\cite{Seznec2005OGEHL}.

\paragraph{Selective Bit Training}
As described, the predictor predicts every lower-order bit in the
target address.  However, many branches have a small set of potential targets.
The probability that a given bit in each target is the same is somewhat high
for these small sets. Thus, the predictor only predicts and trains on a target
bit if that bit differs in the set of potential targets. This selective bit
training reduces the impact of destructive aliasing among weights that are
used for more than one branch.

\paragraph{BTB Compression}
Branch targets in the BTB are represented in a compressed format using the
region-based organization proposed by Seznec for
ITTAGE~\cite{Seznec2011ITTAGE}. There are a number of regions of memory
represented as higher-order target bits and stored in an array. Targets in the
BTB are represented as one of these regions plus an offset.  The number of
regions is small and the number of bits in an offset is less than a full
target, so we can represent a 64-bit target in about half as many bits. The
region array is managed with least-recently-used (LRU) replacement. The BTB
stores only partial tags for branch PCs, trading off the low probability of an
accidental tag match with the storage benefit of partial tags.

\subsection{Implementation}\label{sec:implementation}

The prediction algorithm has several components that may seem challenging
for implementation. Here we give some ideas for balancing complexity and timing.

\paragraph{Computing $y\_{\mbox{out}}$} Perceptron predictors from the
literature and in current processors compute a sum from a vector of weights
read out from distinct tables. BLBP computes $K$ such sums, one per bit of the
target address being predicted. There are $N$ tables whose entries are
$K$-length vectors of weights. The summation operation is carried out in
parallel on each of the $K$ positions across the $N$ tables. In our design,
$N=8$, fewer than the number of tables in recently proposed perceptron
predictors in the literature or industry. Thus, the latency of the computation
is at most the latency of a perceptron-based conditional branch predictor.
The number of adders needed is $N \times K$. In our case, $K=12$ so the
predictor requires 96 small bit width adders for a parallel implementation.  A
mixed-signal implementation of perceptron-based branch predictors can be very
quick and power-efficient even with a large number of inputs~\cite{SNAP}.
Weights are added as analog signals according to Kirchoff's law, and the
transfer function can be implemented by changing the sizes of transistors in
the digital to analog converters.  For a more conventional digital
implementation, the adders are organized as $K$ adder trees with a depth no
more than $\log_2 N$, or 3. The output of the final adders are 8 bits. Thus,
the adders are small.

The parameter $\theta$ is tuned dynamically as described in 
Section~\ref{subsec:snip:optimize}. We find four bits per weight sufficient to 
maintain a good trade-off between accuracy and space-efficiency.

\begin{figure}[htbp!]
  \centering
  \includegraphics[width=0.47\textwidth]{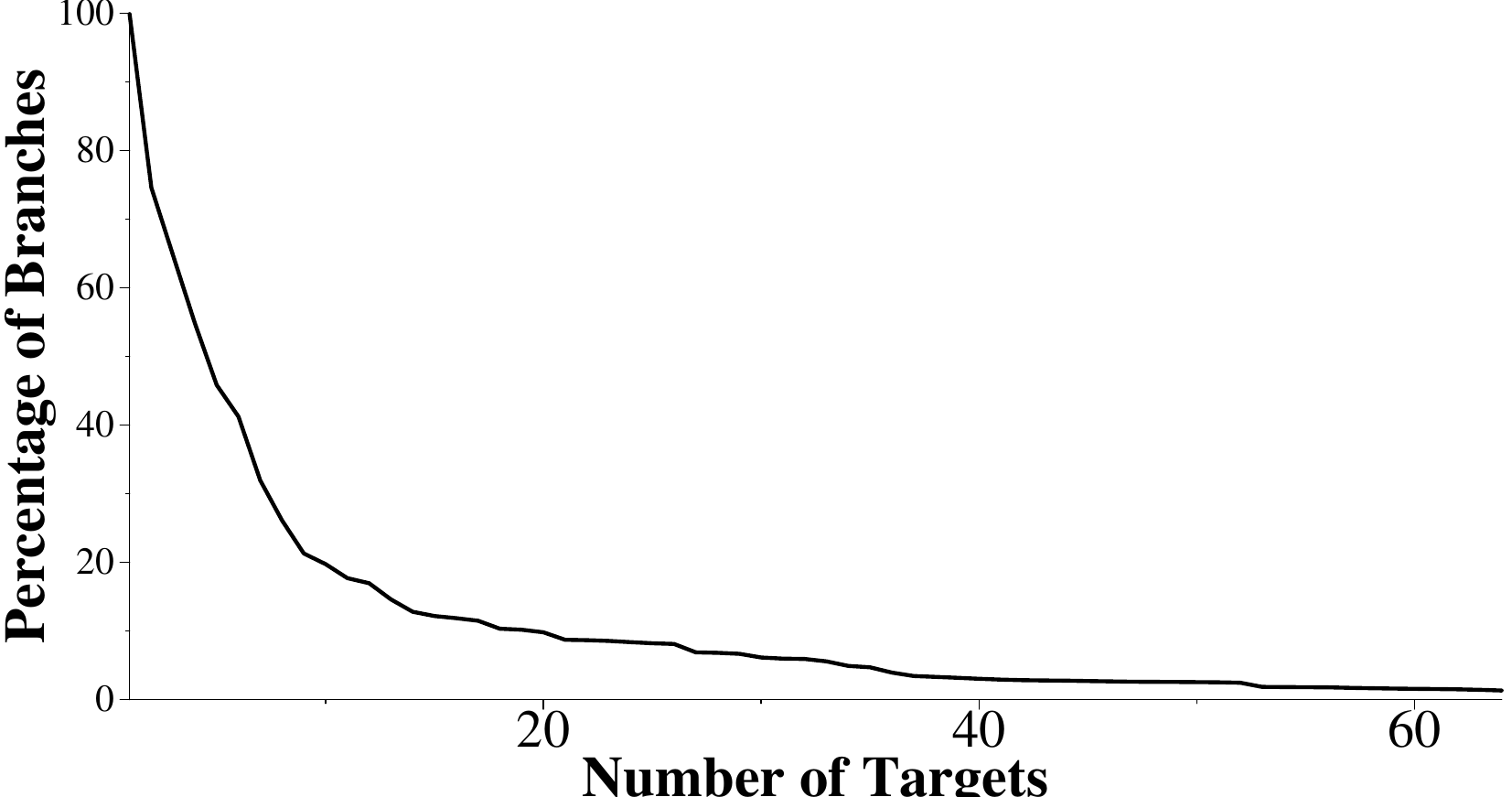}
  \caption{Distribution of Number of Potential Targets}
  \label{fig:targets}
\end{figure}

\paragraph{Computing the Cosine Similarity}
Once the values in $y_{\mbox{out}}$ have settled, the algorithm finds the
non-normalized cosine similarity between $y_{\mbox{out}}$ and the bit vectors
representing the potential target addresses. 

Algorithm~\ref{alg:predict} shows the code on how the BLBP 
prediction scheme computes a vector of $K$ dot product values. Value $K$ 
corresponds with the number of lower-order bits of the target address predicted 
by BLBP. $N$ is the number of different histories. In effect, each of the $N$ 
history lengths gets its own sub-predictor, $M$ is the number of rows in the 
table of perceptron weights, {\em i.e.} the number of possible indices produced by 
the hash of global history.(Thus, indexing value $j$ is bound by $M$.) 
$G_i$ is the history length for the $i$th position in the perceptron predictor, 
determined as geometric history lengths~\cite{Seznec2005OGEHL}. 
$W[N,M,K]$ is a three-dimensional array of 4-bit sign/magnitude integer weights. 
The $W$ array is realized as $N$ SRAM arrays each indexed by a different hash 
of global history to yield a vector of perceptron weights, one weight per address 
bit predicted.

This dot product computation is equivalent to measuring the cosine similarity 
between two vectors, {\em i.e.} the cosine of the angle between two vectors. A 
higher cosine similarity implies a closer match between the two vectors. The 
cosine similarity is usually normalized by the product of the magnitude of the 
two vectors, but in our case this is unnecessary because each bit vector is
multiplied by the same value of $y_{\mbox{\small out}}$, thus the dot product
values are directly comparable.

Computing the non-normalized cosine similarity is simply taking the dot product 
of $y_{\mbox{out}}$ and a bit vector.  This dot product computation can be 
carried out very efficiently by taking the sum of the bitwise AND of each element 
of $y_{\mbox{out}}$ and the corresponding sign-extended bit in the vector. This 
can be accomplished with an adder tree similar to the computation of 
$y_{\mbox{out}}$ itself.

\paragraph{Computing the Cosine Similarity Repeatedly}
There are up to 64 potential target addresses, thus up to 64 possible cosine
similarities to compute.  An associative lookup and parallel cosine similarity
computation would be fast but somewhat expensive in terms of power and
complexity. A sequential search would be more efficient but have higher
latency in cases where there are many targets matching the current branch.
Fortunately, in practice, the vast majority of indirect branches have only a
few observed targets. Figure~\ref{fig:targets} shows the distribution of numbers
of targets.  For each value from 1 to 64 on the $x$-axis, the height of the
curve gives the percentage of all indirect branches having at least that many
distinct targets. We can see from the graph that the majority of indirect
branches have no more than 5 potential targets. Only 10\% of indirect branches
have more than 20 potential targets. Thus, a feasible implementation could
compute 5 cosine similarities per cycle in parallel at a modest cost, taking
only one cycle for over half of all predictions and no more than 4 cycles for
90\% of the predictions. With current decoupled fetch engines, this modest
latency would be hidden for most branches and only exposed in those few cases
with many targets where getting the right prediction matters the most.  Note
that VPC may also incur significant latency in extreme cases with many
potential targets.

\subsection{Searching the BTB}

The algorithm requires collecting all branch targets that match a given branch
PC from the BTB. This can be accomplished with a content addressable memory
(CAM) or by a sequential lookup similar to the strategy taken by the VPC
algorithm~\cite{Kim2007VPC}. Although there are up to 64 potential targets for
a given branch, in practice there are very few targets for each branch and the
sequential strategy can be quick.

    \section{Testing Methodology} \label{sec:meth}

\subsection{Benchmarks}\label{subsec:meth:bench}

\begin{table*}[t]
  \centering
  \begin{tabular}{ |c|c|p{10cm}| } 
   \hline
   \textbf{Benchmark Source}  & \textbf{Number of Benchmarks} & \textbf{Details of Workloads} \\
   \hline
   SPEC CPU2000               & 1                             & \texttt{252.eon} \\
   \hline
   SPEC CPU2006               & 12                            & \texttt{400.perlbench}, \texttt{403.gcc}, \texttt{453.povray}, \texttt{458.sjeng} \\
   \hline
   SPEC CPU2017               & 7                             & \texttt{600.perlbench}, \texttt{602.gcc}, \texttt{623.xalancbmk} \\
   \hline
   CBP-5 Competition          & 68                            & Industry-sourced workloads offered for CBP-5 by host Samsung. Benchmarks are divided into two categories: \texttt{MOBILE} and \texttt{SERVER}. Trace lengths also fall in two categories: \texttt{SHORT} and \texttt{LONG}. \\ 
   \hline
  \end{tabular}
  \caption{Description of the 88 workloads used for testing and evaluation. The benchmarks come from four sources, including a Championship Branch Prediction competition.}
  \label{tab:bench}
\end{table*}

For evaluation, we use a suite of 88 workloads gathered from
multiple benchmark suite sources.  Represented are SPEC
CPU2000~\cite{SPEC2000}, SPEC CPU2006~\cite{Henning2006SPEC} and SPEC
CPU2017~\cite{SPEC2017} benchmark suites and traces from most recent JILP
Branch Prediction Competition~\cite{CBP5website} held at ISCA 2016.

Traces were selected if they averaged more than one misprediction per 1000
instructions for an initial simulation using an infinite-sized branch target
buffer.  Table~\ref{tab:bench} summarizes the sources and qualities of our
benchmarks.

The Samsung-sourced benchmark \texttt{LONG-MOBILE-8} contains \textit{more}
indirect branches than conditional.  In general, the Samsung workloads tend to
have more indirect branches compared to the other workload sources.  Although
Samsung did not provide much information about the benchmarks used, it is
reasonable to expect that many of the workloads come from Android workloads in
the Java language.  Since Java relies on virtual method dispatch by default,
it is not surprising to find many indirect branches in these traces.

\subsection{Simulation Setup}\label{subsec:meth:setup}

\begin{table*}
  \centering
 \begin{tabular}{ |c|p{9cm}|c| } 
   \hline
   Indirect Branch Predictor & Implementation Configuration                     & Total Hardware Budget \\
   \hline
   BTB        & 32K-entry, partially-tagged, direct-mapped branch target buffer & 64 KB  \\
   \hline
   VPC        & 32K-entry, partially-tagged, direct-mapped BTB with \newline Multi-perspective 
                Perceptron Predictor for conditional branch prediction          & 128 KB  \\
   \hline
   ITTAGE     &                                                                 & 64 KB \\
   \hline
   BLBP       & 64-entry, 64-way set-associative, partially-tagged IBTB,\newline 
                256 10-bit local histories, 630-bit global history, \newline
                8 correlating-weights tables, and 128-entry region array        & 64.08 KB \\
   \hline
  \end{tabular}
  \caption{The implementation setups for the indirect branch predictors evaluated.}
  \label{tab:branch}
\end{table*}

We use the branch prediction simulation infrastructure released for the
Championship Branch Prediction competition~\cite{CBP5website}.  The simulation
infrastructure has been augmented with additional code for a BTB and an
assortment of indirect branch predictors including BLBP.  For conditional
branch prediction, we use a hashed perceptron
predictor~\cite{Tarjan2005Hashed}.

The appropriate metric for branch prediction studies is mispredictions per
kilo-instruction (MPKI).  Unlike simple misprediction rate, MPKI takes into
account the relative frequency of branches compared to other instructions.
Previous work has demonstrated a linear relationship between MPKI and
performance~\cite{zheiiswc}.  Thus, it is sufficient to measure MPKI to infer
an impact on performance.

Table~\ref{tab:branch} gives detailed information on the indirect branch
predictors implemented for this study.  We use a BTB as our indirect branch
prediction baseline.  The baseline BTB is a 32K-entry cache indexed by branch
address and filled with the most recently observed branch target for that
branch address.  This BTB is large compared with recent examples from
industry.  For example, the BTB in Samsung's recent Mongoose processor has 4,096
entries~\cite{Mongoose}.  Thus, this BTB provides a practical upper limit on
the accuracy of modern BTBs.

We implement the Virtual Program Counter (VPC) predictor from Kim {\em et
al.}~\cite{Kim2007VPC} using a 64KB Multi-perspective Perceptron Predictor
(MPP)~\cite{JimenezCBP5} as its underlying conditional branch predictor.  The
original VPC implementation relies on global branch history to make
predictions for indirect branches.  However, our implementation is a hashed
perceptron predictor that uses a set of 37 features to provide alternate
perspectives on branch history.  Our implementation of VPC has a 32K-entry BTB
for target storage.  This modified VPC gave an MPKI of 0.29 for indirect
branches while maintaining a low degradation of 2.05\% in the prediction
accuracy of conditional branches.

Finally, we set up BLBP as described in Section~\ref{sec:snip}.  There are 8
independently accessed SRAM arrays for 8 different history intervals.  Each
SRAM array has rows of 12 4-bit vectors implementing the perceptron weights.
There is a 630-bit global history and a table of 256 10-bit local histories.
There is an indirect branch target buffer (IBTB) consisting of 64 sets of 64
entries, each of which contains an 8-bit partial tag, a 7-bit region number, a
20-bit region offset, and a 2-bit re-reference interval prediction~\cite{RRIP}
for replacement.  The region number and descriptor allow a compressed
representation of targets in the same manner as the ITTAGE
predictor~\cite{Seznec2011ITTAGE}.  The total state for the prediction tables,
histories, IBTB, and region table is approximately 64KB, allowing an iso-area
comparison with the ITTAGE implementation from the second branch prediction
competition.

Each of the SPEC traces is a simpoint~\cite{SimPoint} of one billion
instructions.  The Samsung-sourced traces are of variable length averaging
in the 100s of millions of instructions.  For our experiments we measure
branch target MPKI over each entire trace.

    \section{Results} \label{sec:results}

\subsection{Overall Performance}\label{subsec:results:overall}

\begin{figure*}[htbp!]
  \centering
  \includegraphics[width=1.0\textwidth]{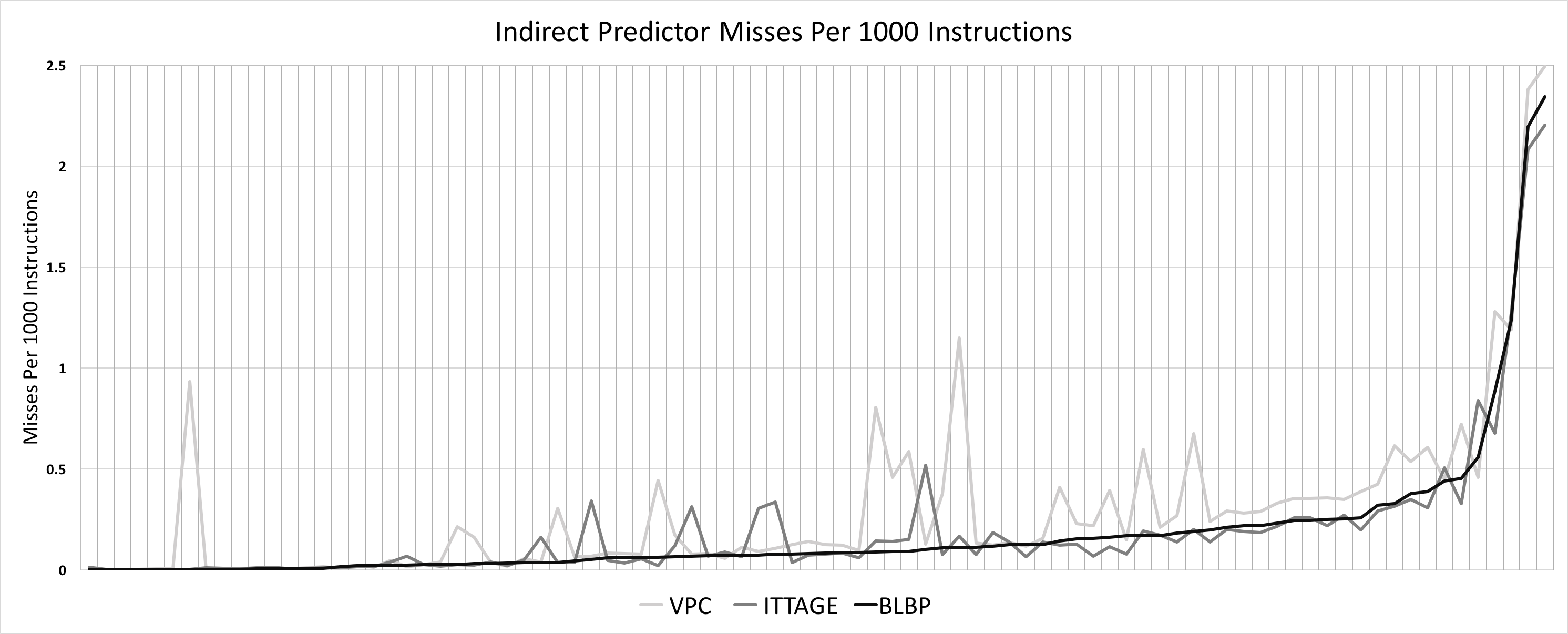}
  \caption{MPKI for 3 of 4 indirect branch predictors on a suite of 88 higher-MPKI benchmarks. BTB MPKI has been omitted due to much-higher MPKI relative to the other predictors. Benchmarks are sorted by BLBP MPKI.}
  \label{fig:mpkis}
\end{figure*}

\begin{figure*}[htbp!]
  \centering
  \includegraphics[width=1.0\textwidth]{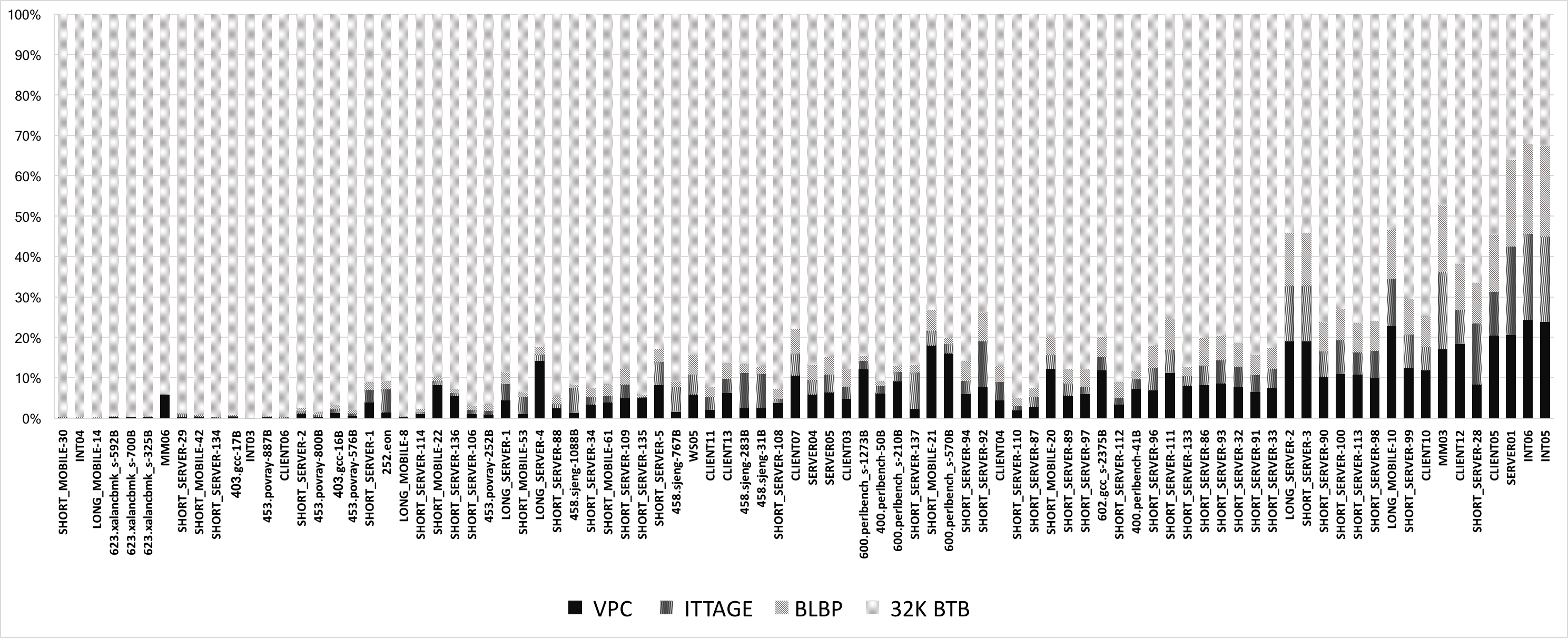}
  \caption{Percentage breakdown of predictor MPKI performance relative to each other. Benchmarks are sorted by BLBP MPKI.}
  \label{fig:percs}
\end{figure*}

\begin{figure*}[htbp!]
  \centering
  \includegraphics[width=1.0\textwidth]{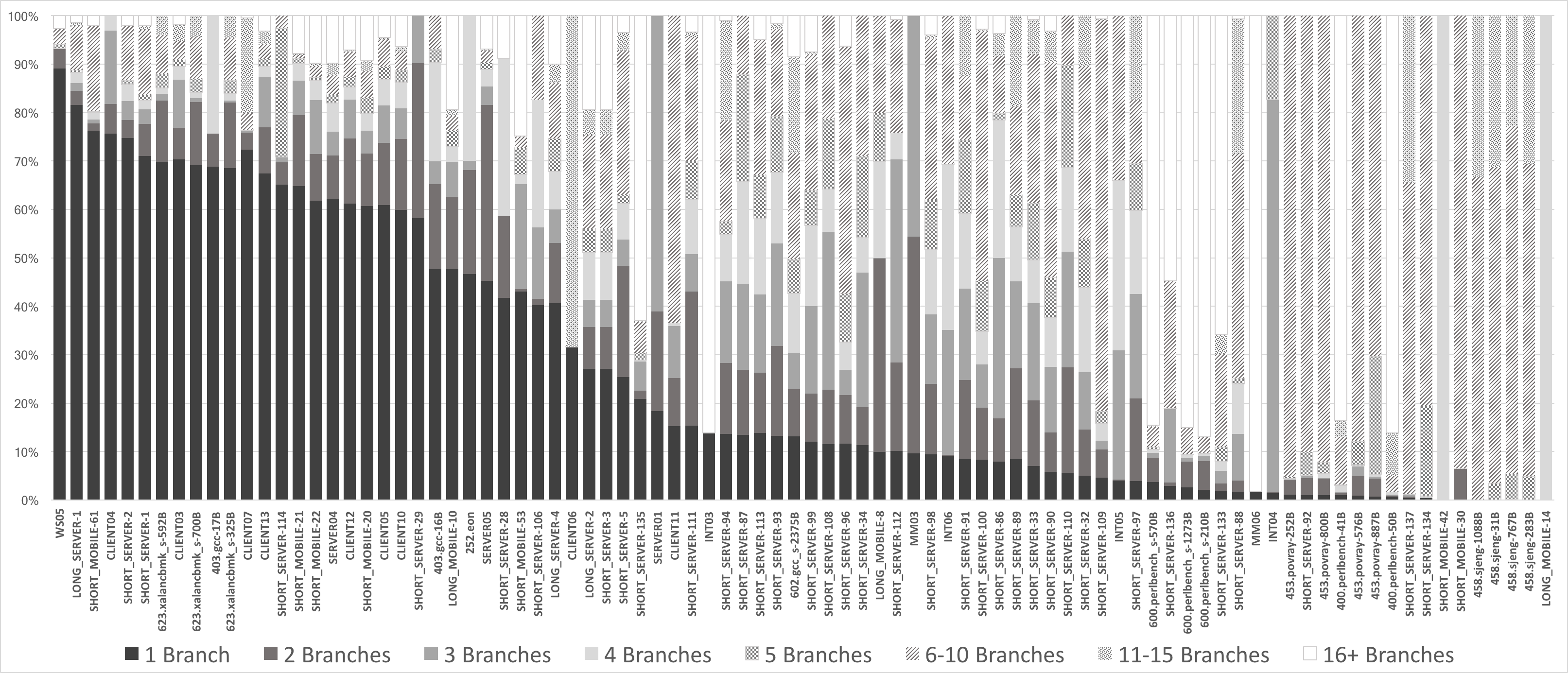}
  \caption{Monomorphism and polymorphism in indirect branches can be observed with the number of targets each branch instruction accesses. Samsung trace \texttt{WS05} is shown to be highly monomorphic, with 90\% of its indirect branches accessing only one target value. Traces are ordered from most monomorphic to most polymorphic, relative to the percentage of instructions with more than one target address.}
  \label{fig:count}
\end{figure*}

Among the four predictor implementations, the baseline BTB predictor performed
the worst, with an arithmetic mean of 3.40 MPKI across the benchmarks.  Next
was VPC with a mean of 0.29 MPKI.  ITTAGE and BLBP rounded out our results
with 0.193 and 0.183 MPKI, respectively.  Thus, for the benchmark suite
tested, BLBP led to a 5\% improvement in MPKI over the state-of-the-art.

In response to reviwer concerns, we also tested the predictors with the CBP4~\cite{CBP4website}
workloads, which are different from the CBP5 traces used to develop BLBP.
Without tuning either predictor, ITTAGE yields 0.028 MPKI while BLBP achieves
0.027 MPKI, outperforming ITTAGE by 3.5\%.

Figure~\ref{fig:mpkis} presents the MPKI results for three of the four 
predictors, with the BTB baseline being omitted due to its very high MPKI 
relative to the other predictors. The benchmarks (unlabeled) are sorted by 
monotonically increasing BLBP MPKI performance.
    
Figure~\ref{fig:percs} shows similar data, but reveals MPKIs of the four 
predictors compared to each other.

\subsection{Effect of Optimizations}

\begin{figure}[htbp!]
  \centering
  \includegraphics[width=0.47\textwidth]{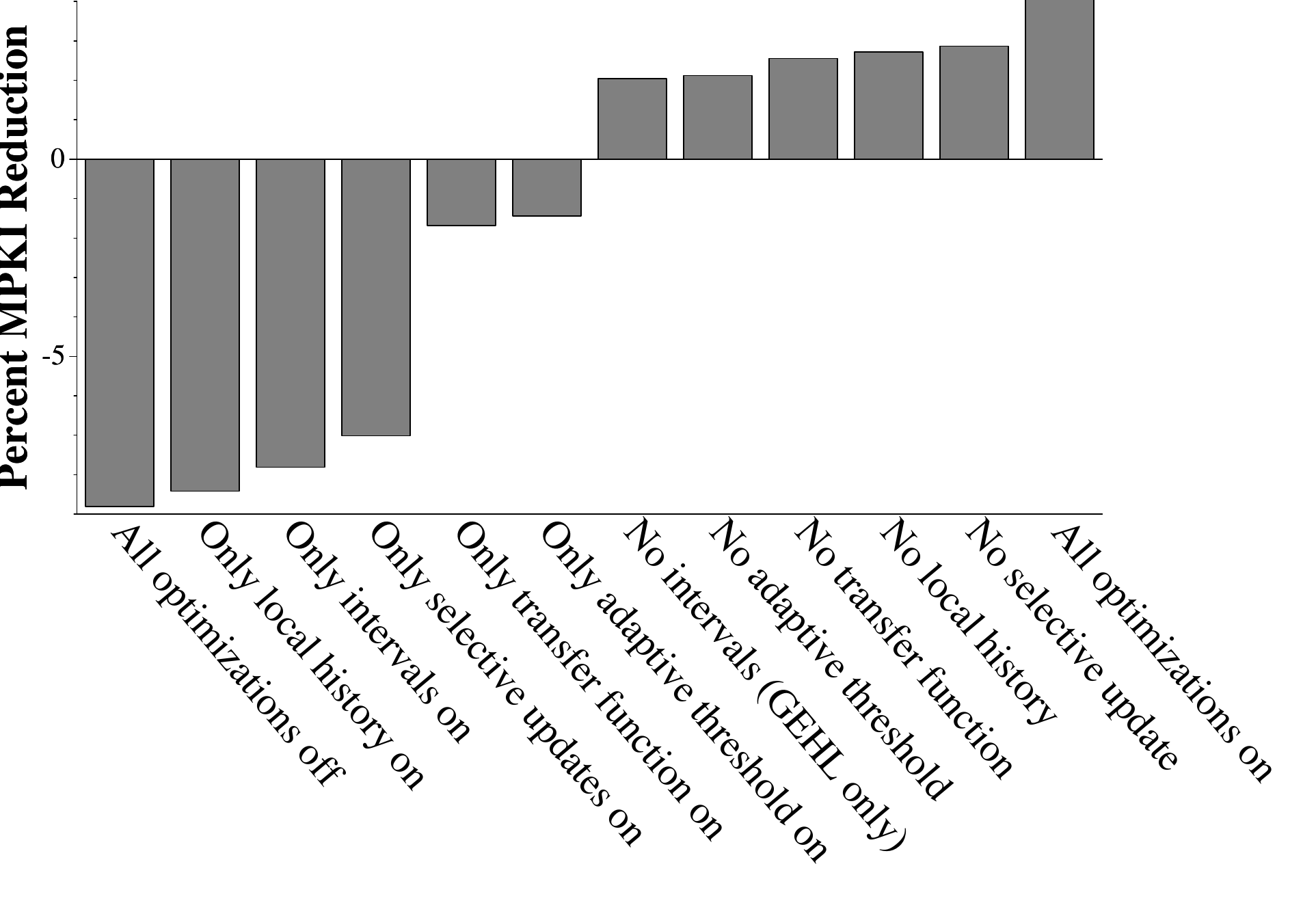}
  \caption{Effect of Optimizations}
  \label{fig:optimizations}
\end{figure}

Figure~\ref{fig:optimizations} illustrates the effect of the optimizations
applied to BLBP. The figure shows the percent improvement in average MPKI over
ITTAGE for versions of the predictor for configurations where only one
optimization is enabled, one of the optimizations is disabled, and all or none
of the optimizations are enabled. With all optimizations turned on, BLBP
achieves 5.3\% improvement over ITTAGE.  With no optimizations, BLBP achieves
an average 8.8\% higher MPKI than ITTAGE.  Thus, doing at least some
optimizations is essential to BLBP.

Let us explore what happens when turning on only one of the optimizations at a
time. Turning on the local history improves the predictor the least, followed
by using intervals instead of geometric history lengths (GEHLs), and
selectively updating only bits that differ in targets. Using a non-linear
transfer function provides a large boost to the otherwise unoptimized
predictor, bringing it within 1.7\% of ITTAGE. Adaptive training provides the
largest boost, only 1.4\% higher MPKI than ITTAGE.

Now let us see what happens when turning off only one optimization, keeping
the rest.  The biggest change occurs when using GEHL histories instead of
intervals; the improvement over ITTAGE drops from 5.3\% to 2.04\%. Thus, in
concert with the other features, intervals seem to provide an important boost.
Omitting selective bit updates seems to hurt the least, lowering MPKI
improvement to 2.86\%. However, omitting any of the optimizations produces a
significant reduction in MPKI improvement. Thus, the optimizations seem to
work synergistically (if not additively) to improve accuracy.

\subsection{Effect of Associativity}

\begin{figure}[htbp!]
  \centering
  \includegraphics[width=3in]{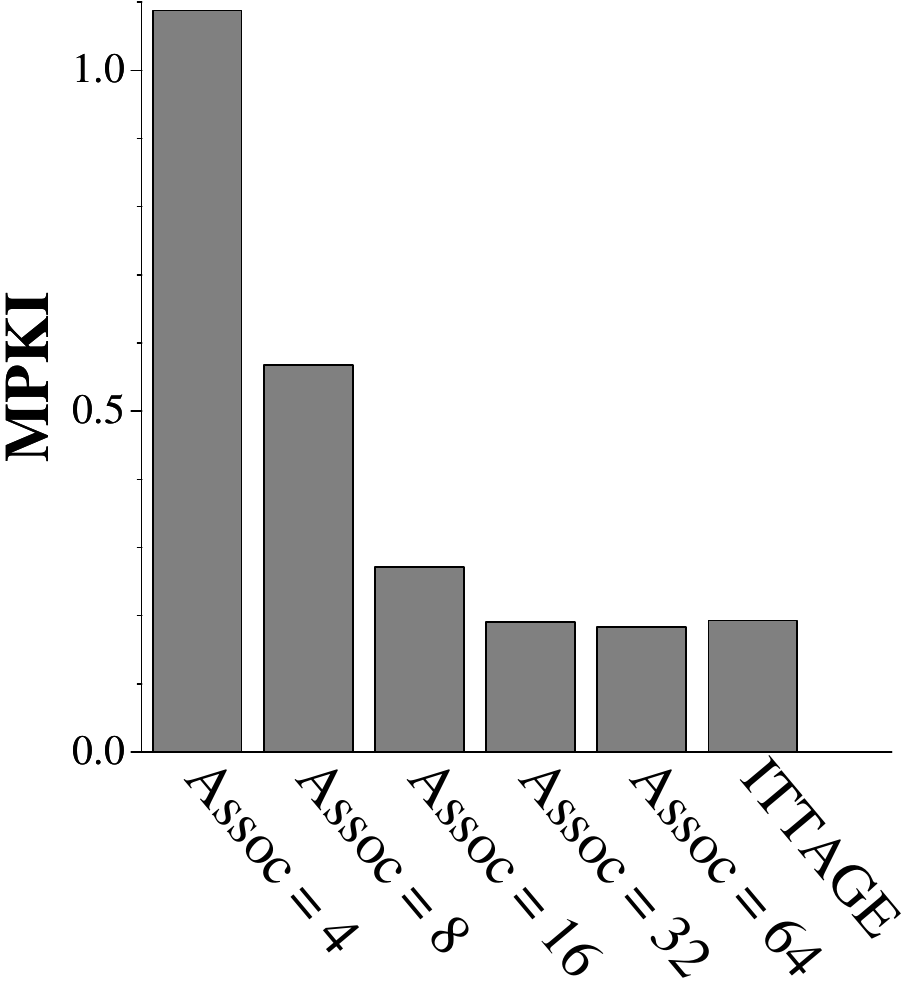}
  \caption{Effect of Associativity}
  \label{fig:assoc}
\end{figure}

The IBTB is a set-associative structure. Each set stores 64 targets observed
from indirect branches that hash to that set. Although, as we have seen in
Section~\ref{sec:implementation}, most branches have at most 5 targets, many
branches may hash to the same IBTB set causing collisions, and some branches
actually require many targets. Thus, the associativity of the structure must
be large enough to combat conflict misses and to accommodate branches with
many targets. Figure~\ref{fig:assoc} shows the effect on accuracy of varying
the associativity of the IBTB, keeping the number of IBTB entries the same at
4,096. A 4-way IBTB gives an unacceptably high 1.09 MPKI. An 8-way IBTB cuts
the MPKI almost in half to 0.57. A 16-way BTB gives another large improvement,
down to 0.27 MPKI. With 32-way set associativity, the IBTB yields an MPKI of
0.19, equivalent to ITTAGE.  The 64-way IBTB gives 0.183 MPKI, a 5\%
improvement over ITTAGE.  Thus, clearly high associativity is necessary for
good accuracy.

    \section{Conclusion \& Future Work} \label{sec:conc}
In this paper, we introduced Bit-Level Perceptron-Based Indirect Branch
Predictor, or BLBP.  BLBP predicts indirect branch targets' select lower-order
bits using perceptron-based learning.  A selection of known target addresses
for the branch are gathered from the IBTB for comparison; the target address
that matches closest with the predicted bits is output as BLBP's prediction.

We have shown BLBP to outperform the state-of-the-art. Using a suite of 88
benchmarks with significant indirect branches, we show BLBP improves upon
ITTAGE's prediction performance by 5\%, reducing MPKI from 0.193 to 0.183.

In future work we plan to improve accuracy by exploring features beyond global
and per-branch history. For example, the recently proposed multiperspective
perceptron predictor uses a variety of control-flow features to improve
accuracy~\cite{JimenezCBP5}. We plan to reduce the complexity of the
computations involved. The original perceptron predictor was quite complex
with a high latency, but follow-up work reduced the latency through
ahead-pipelining and other
techniques~\cite{Jimenez2003Path,Jimenez2005Piecewise,LohJimenezIJPP}. We are
confident that the same sorts of techniques can be applied to BLBP to achieve
lower latency and power while maintaining high accuracy. We plan to explore
ways of avoiding the high-associativity of the IBTB, perhaps using a hierarchy
of structures~\cite{Driesen1998Cascaded,JimenezMICRO2000}.  We also plan to
explore how BLBP might be used to predict conditional branches as well as
indirect branches as VPC does, allowing consolidation of the two structures.

\section{Acknowledgments}

We thank the anonymous reviewers for their valuable comments and feedback to
improve the content and quality of this paper.  We also thank the National
Science Foundation, which supports this work through grant CCF-1649242.
Portions of this research were conducted with the advanced computing resources
provided by Texas A\&M High Performance Research Computing.

\end{document}